\documentclass[11pt]{article}

\usepackage{epsfig,graphics}

\usepackage[english,francais]{babel}

\def\be{\begin{equation}}
\def\ee{\end{equation}}
\def\benn{$$}
\def\eenn{$$}
\def\bea{\begin{eqnarray}}
\def\eea{\end{eqnarray}}
\def\beann{\begin{eqnarray*}}
\def\eeann{\end{eqnarray*}}

\def\eqdef{\stackrel{\mbox{\tiny def}}{=}}     
\newcommand{\ket}[1]{|\kern.3ex#1\kern.3ex\rangle}
\newcommand{\bra}[1]{\langle\kern.3ex #1 \kern.3ex|}
\newcommand{\APPROX}[1]{                
   {{\raisebox{-.3cm}{$\textstyle\simeq$}} \atop {\scriptstyle{#1}}}}
\newcommand{\APPROXs}[1]{               
   {{\raisebox{-.3cm}{$\textstyle\sim$}} \atop {\scriptstyle{#1}}}}
\newcommand{\mean}[1]{\left\langle #1 \right\rangle} 

\newcommand{\EXP}[1]{{\mbox{\large e}}^{#1}}         
\newcommand{\re}{\mathop{\mathrm{Re}}\nolimits}      
\newcommand{\im}{\mathop{\mathrm{Im}}\nolimits}      
\newcommand{\tr}[1]{\mathop{\mathrm{Tr}}\nolimits\left\{ #1 \right\}}  
\newcommand{\cotg}{\mathop{\mathrm{cotg}}\nolimits}  
\newcommand{\sign}{\mathop{\mathrm{sign}}\nolimits}  

\def\I{{\rm i}}                     

\newcommand\dz{\partial_z}
\newcommand\dzb{\partial_{\bar z}}
\newcommand{\drond}[2]{\frac{\partial #1}{\partial #2}} 

\def\L{{\rm L}}                  
\newcommand\oc{\omega_c}
\newcommand\moc{\langle\omega_c\rangle}

\def\plm{\Pi^\L_-}               
\def\plp{\Pi^\L_+}
\def\blm{{\rm B}^\L_-}
\def\blp{{\rm B}^\L_+}

\def\plmm{\Pi^{\mean{\L}}_-}
\def\plpm{\Pi^{\mean{\L}}_+}

\begin{document}

\selectlanguage{english}

\title{Hall Conductivity in the presence of repulsive magnetic impurities}

\author{ Jean Desbois,         St\'ephane Ouvry and  
         Christophe Texier}
\date{}

\maketitle
\noindent
{Laboratoire de Physique Th\'eorique et Mod\`eles Statistiques,
 91406 Orsay C\'edex, France.}

\begin{abstract}
The 
Hall conductivity of disordered magnetic systems consisting of
hard-core point vortices randomly 
dropped  on the plane with a Poissonian distribution, 
has a behavior analogous to
the one 
observed experimentally by R.~J.~Haug, R.~R.~Gerhardts, K.~v.~Klitzling and
K.~Ploog, with repulsive scatterers \cite {1}. 
We also argue that  models of homogeneous magnetic field with 
disordered potential,
have necessarily
vanishing Hall conductivities when their
Hilbert space is restricted to a given Landau level subspace.

\end{abstract}


It is commonly believed \cite{0} that in quantum Hall devices disorder plays
a crucial role in the understanding of plateaus for the Hall conductivity as
a function of $1/B$ or $N(E_F)$, the number of electrons, at integer (or
fractional) values in units of $e^2/h$. For an homogeneous $B$ field,
 the linear response of the system to a small electric field gives no hint
 of such a remarkable behavior since all the states are delocalized and have
 the same transverse conductivity which varies linearly with $1/B$ or $N(E_F)$ 
 (classical straight line). Disorder is needed to explain why some states
 (in fact most of them) are localized in broadened Landau levels, thus the
 plateaus in the Hall conductivity, {\bf around} the classical line.  
 However,  R. J. Haug et al \cite{1}
 reported the experimental observation of a {\bf shifted} quantized
Hall conductivity with respect to the classical line, when attractive or 
 repulsive scaterrers are considered in the Hall sample.
The data - Hall conductivity versus the filling factor- are
shifted to the left when the scaterrers are repulsive, and to the right when
they are attractive. The authors in \cite{1} were able to reproduce 
 qualitatively these shifts within a self-consistent T-matrix approximation 
  computation.
 They  also argued  that two-dimensional disordered 
$\delta$ repulsive models  projected in the
lowest Landau level (LLL) of an external magnetic field \cite{2}, which exhibit
an asymmetrically
disorder induced broadened LLL, indeed favour this phenomenon. The 
root of the understanding lies in the fact that
 conducting states are not  located anymore at the center of a
symmetrically broadened Landau level, as it is the case for neutral
scatterers, but in one side or the other of the now asymmetric density of
states (DOS). On the theoretical side on the other hand, the question was
asked  in \cite{2} about the 
way to compute exactly the average LLL Hall conductivity.

In the following, we  study a model where the disorder is contained 
in the definition of the magnetic field itself. We 
 consider \cite{3} a gas of electrons coupled to
  hard-core point vortices, hereafter called magnetic 
impurities, carrying a
flux $\phi=\alpha\phi_o$ ($\phi_o=h/e$ is the quantum flux) and randomly
dropped on the plane according to a Poisson distribution, with Hamiltonian
(in units $m_e=\hbar=1$)
\be\label{Hpauli}
H=\frac{1}{2}\left(\vec p-e\vec A(\vec r)\right)^2
-\sigma_z\frac{e}{2}B(\vec r)
\ee
The Aharonov-Bohm vector potential 
$e\vec A(\vec r)=\alpha\sum_i 
\frac{\vec k\times(\vec r-\vec r_i)}{|\vec r-\vec r_i|^2}$ and the magnetic
field
$\vec B(\vec r)=\phi \sum_i\delta(\vec r-\vec r_i)$
depend on the configuration of the random positions $\vec r_i$
of the impurities. It is always possible to take $\alpha\in[0,1/2]$
because of the periodicity of period 1 in $\alpha$ and of the symmetry
$\alpha\to-\alpha$.
The spin assignation in $(1)$ is needed to define in a non ambiguous way 
 the model at hand. We choose $\sigma_z=-1$ which yields a  short
distance regularization for  the impurities such that  they are hard-core.
It was shown in \cite{3} that
two distinct  spectral behaviors for the average DOS occur when $\alpha$
varies from $\alpha=1/2$ (i.e. big flux) and $\alpha\simeq 0$.  When
$\alpha$ is big, the electrons see the
impurities individually, a maximum disordered situation with
a free DOS but   a depletion
of states at origin of the spectrum. In contrast, for small value of $\alpha$,
the inhomogeneities of the disordered
magnetic field are less relevant, therefore a Landau like average DOS,
with Landau oscillations, i.e. Landau levels
separated by a mean Landau gap $e\mean{B}$ and
broadened by disorder ($\mean{B}=\rho\phi$ is the mean magnetic
field through the
plane,  $\rho$ is the mean  impurity density -
if $\rho$ is taken to be of the order of the density of current carriers
$\rho=4. 10^{15}m^{-2}$, one obtains, for $\alpha=1/2$,
a  mean magnetic field   precisely in the
experimental range of the Quantum Hall Effect
$ \mean{B}\simeq 10T$
).
Thus, in the  small $\alpha$ limit, the random magnetic impurity model
has the required properties, i.e. an average magnetic field with disorder
induced broadened Landau levels,
to induce localization and eventually  a quantized Hall conductivity.

The  non unitary transformation ($\moc =e\mean{B}/2$)
\be\label{transf}
\psi=\EXP{-\frac{1}{2}\moc r^2}
\prod_{i=1}^N |\vec r-\vec r_i|^\alpha \:\tilde\psi'
\ee
leads to the equivalent Hamiltonian 
\be\label{Htilde'IM}
\tilde{H}'=\frac{1}{2}\plpm\plmm
- \I\alpha\left( \Omega -\mean{\Omega} \right) \plmm
\ee
where $\Omega=\sum_i\frac{1}{\bar z-\bar z_i}$, and $\plpm$ and $\plmm$
are the covariant Landau operators for the mean
magnetic field. The Hamiltonian
(\ref{Htilde'IM}) has the simple structure of a Landau Hamiltonian for the mean
magnetic field plus a disordered potential. One might consider
that the Hall conductivity computation could be
simplified if 
the mean magnetic field is sufficiently strong so that
one can neglect couplings between
Landau levels, or, more drastically, retain only the LLL. Remarkably enough 
\cite{5}, the Hamiltonian (\ref{Htilde'IM}), when projected in the LLL
of the mean magnetic field,  precisely yields the repulsive 
$\delta$ impurity Hamiltonian \cite{2} in the LLL of the mean magnetic field 
\be\label{LLL} 
H=H_{\rm \mean{LLL}} + \lambda \sum\delta(\vec r-\vec r_i)
\ee
with $\lambda=2\pi\alpha$. So the question:
What is the  average Hall conductivity for the LLL Hamiltonian (\ref{LLL})
and, more generally, for  Hamiltonians
of the type  $H=H_{L}+V(\vec r)$, where $V(\vec r)$ is a disordered potential,
when $H$ is restricted to the Hilbert space of a given Landau level
 of $H_{L}$?
 The answer is: 
 In the linear response formalism, 
 such  a conductivity vanishes, implying that a non
vanishing conductivity necessarily  arises from couplings between 
different Landau levels. We insist here that restricting the Hilbert space
to a given landau level and computing the conductivity in this given subspace
should not be confused
with the problem of evaluating the contribution of a given Landau level to
the total conductivity.
The Hamiltonian restricted to the $n$th Landau level is
$H^{(n)}=E_n P_n + P_n V P_n$
with $E_n=2\oc(n+1/2)$ and $P_n$
the energy and the
projection operator of the $n$th Landau level
$
P_n(z,z')\eqdef\bra{z}P_n\ket{z'}
=\frac{\oc}{\pi} L_n(\oc|z-z'|^2)\,
\EXP{-\frac{1}{2}\oc(|z-z'|^2 - z\bar z' + \bar z z')}
$
and
$\plp P_n=(1-\delta_{n,0}) P_{n-1}\plp$ (accordingly
$P_n \plm=(1-\delta_{n,0}) \plm P_{n-1}$).
The complex thermalized conductivity for one electron  
$\sigma^-_\beta(t)\equiv \sigma_{xx}(t)-i\sigma_{yx}(t) $  rewrites as \cite{4}
\be\label{oui}
\sigma^{-(n)}_\beta(t)=\theta(t)\frac{\I e^2}{2VZ^{(n)}_\beta}
\int dz d\bar z dz' d\bar z'\,
\left(G^{(n)}_{\beta-\I t}(z',z)\, \Pi_-^{L} G^{(n)}_{\I t}(z,z') \,z'
     - (\I t\to\beta+\I t) \right)
\ee
where the   propagator
$G^{(n)}_{\beta}(z,z')$ for 
$H^{(n)}$ is by definition
\be 
G^{(n)}_{\beta}(z,z')=\bra{z}P_{n}e^{-\beta H^{(n)}}\ket{z'}.
\ee
The operator $\plm$ in (\ref{oui}) happens to be 
flanked by two projectors. Since $P_n\plm P_n=0$, 
then necessarily $\sigma^{-(n)}_\beta(t)=0$, implying that 
the Hall conductivity for a gas of non interacting electrons vanishes as 
well when restricted to a given  Landau level.

It follows that the full Hamiltonian (\ref{Htilde'IM}) is needed to get a non trivial information on the
conductivity. First order perturbation theory  (Fig.1) gives a
behavior for the Hall conductivity \cite{4}
 which is quite reminiscent of the experimental data in \cite{1} for
repulsive scatterers.
Of course we do not pretend to describe precisely this particular
experimental situation.
Still, we would  like to emphasize that
an enhancement of the Hall conductivity does appear in the presence
of repulsive magnetic impurities, and that this phenomenon can be obtained 
only if all  Landau levels are considered.

Acknowledgments: S.O. would like to thank J. Bellissard for indicating
\cite{1}.

\begin{figure}[h]
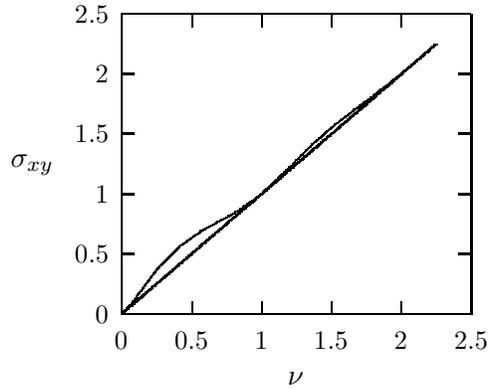

\begin{center}

\setlength{\unitlength}{0.240900pt}
\ifx\plotpoint\undefined\newsavebox{\plotpoint}\fi
\sbox{\plotpoint}{\rule[-0.200pt]{0.400pt}{0.400pt}}%


\caption{\label{syx} Hall conductivity in unit of $e^2/h$
of the random magnetic impurity model at first order in $\alpha$
for $\alpha=0.01$ as a function of the filling factor 
$\nu=\frac{Nh}{Ve\mean{B}}$
; straight line = classical result, full  line = perturbative result}
\end{center}
\end{figure}

\end{document}